# *The Adoption of Artificial Intelligence in Different Network Security Concepts*

Mamoon A. Al Jbaar
Department Computer and Information Engineering
College of Electronics Engineering, Ninevah University
Iraq, Mosul
mamoon.thanoon@uoninevah.edu.iq

Adel Jalal Yousif
Electronic Computing Center
University of Diyala
Diyala, Iraq
adil.jalal.yousif@uodiyala.edu.iq

Qutaiba I. Ali
University of Mosul
Iraq

## Abstract

The obstacles of each security system combined with the increase of cyber-attacks, negatively affect the effectiveness of network security management and rise the activities to be taken by the security staff and network administrators. So, there is a growing need for the automated auditing and intelligent reporting strategies for reliable network security with as less model complexity as possible. Newly, artificial intelligence has been effectively applied to various network security issues, and numerous studies have been conducted that utilize various artificial intelligence techniques for the purposes of encryption and secure communication, in addition to using artificial intelligence to perform a large number of data encryption operations in record time. The aim of the study is to present and discuss the most prominent methods of artificial intelligence recently used in the field of network security including user authentication, Key exchanging, encryption/decryption, data integrity and intrusion detection system.

# 1. INTRODUCTION

With the continuous development and growth usages of Internet and networking, network security has become one of the most important research topic. Network security consists of the procedures, strategies, and policies that are implemented to monitor, detect, prevent and control unauthorized access including abuse, alteration, or denial of network reachable resources. The key challenge in network security field is how to identify and handle different types of network malicious attacks especially those attacks never seen before for which there is no prior information to deal with. Conventional network security depends mostly on static monitoring of security devices distributed on specific nodes or edges such as firewalls, protocols, anti-virus software, intrusion detection and prevention systems which are controlled by the administrator of network [1].

However, traditional network security techniques have many restrictions for providing distinctive feature descriptors to describe network problems related to malicious attacks, due to their limitations in model complexity [2]. Additionally, cyber attackers are becoming more destructive and sophisticated utilizing zero-day exploits, advanced persistent threats (APTs) and malware that avoid security measures and can operate for long periods of time without being detected [3]. Recently, the increased interest in the artificial intelligence approach resulted in essential advancements of pattern recognition or anomaly detection mechanisms that are employed successfully in network security issues. Artificial intelligence provides many techniques, most of which are inspired by nature computational methods such as neural networks, genetic algorithm, fuzzy logic machine learning, data mining and others [4].

## 2. Artificial Intelligence Concepts

The term artificial intelligence (AI) is refers to the science and engineering of making intelligent machines that are programmed to learn and think like humans beings and emulate their behaviors mainly based on machine learning algorithms. While, machine learning is a branch of artificial intelligence that makes models learn and improve themselves from experience automatically without explicit human intervention [5]. Machine learning algorithms generally are trained to make prediction or classification task. Learning algorithms of artificial intelligence can be categorized into three broad classes depending on the way in which they learn about data for decision making as: Supervised learning, Unsupervised learning and Reinforcement learning [3].

Supervised learning is the machine learning task in which a model learns mapping an input to an output depending on an input - desired output labels. As input data is loaded into the model, it modifies its weights until the model has been fitted appropriately based on the given examples.

Unsupervised learning is a kind of machine learning where a model must look for patterns in a dataset with no labels and with minimal human supervision. This is in contrast to supervised learning techniques, such as classification or regression, where a model is given a training set of inputs and a set of observations, and must learn a mapping from the inputs to the observations. In unsupervised learning, only the inputs are available, and a model must look for interesting patterns in the data .

Reinforcement learning is the science of decision making. It is about learning the optimal behavior in an environment to obtain maximum reward. This optimal behavior is learned through interactions with the environment and observations of how it responds, similar to children exploring the world around them and learning the actions that help them achieve a goal. Various approaches in the AI domain has been presented in this work to automate the task of network security while decreasing human factor as explained below [6].

## 2.1 Artificial Neural Network (ANN)

An artificial neural network is a portion of a computing system that simulates how the human brain analyzes and processes data. It is the cornerstone of artificial intelligence and is capable of resolving problems that would be considered impossible or difficult to solve by human or statistical criteria. Self-learning characteristics enable ANNs to improve their performance as new data becomes available.

## 2.2 Deep Neural Network (DNN)

A deep neural network is a type of neural network that has a particular level of complexity, typically more than two layers. Deep neural networks analyze data in complicated ways through the use of advanced mathematical modeling. it offers a lot of value to statisticians, particularly in increasing accuracy of a machine learning model.

## 2.3 Convolutional Neural Networks (CNNs)

Convolutional neural networks are a type of artificial neural network that is widely used to tackle based on computer vision such as image recognition and categorization. As in the case with every neural network, it solves the problem by determining the best approximation to a function. This is accomplished through the use of weights and bias that are changed through the use of a back-propagation algorithm. CNNs require minimal pre-processing for preparing data input .

## 2.4 Recurrent Neural Network (RRN)

Recurrent neural network is a type of neural network that has loops that enable the network to store information. Generally, recurrent neural networks use their prior experiences based on Long Short Term Memory (LSTM) strategy to adjust matrix weight of its nodes. Recurrent models are very advantageous because they can handle sequence of vectors, which allows for more complex tasks to be performed effectively.

## 3.Network security Studies and Artificial Intelligence

Constructing an intelligent algorithms for encryption/decryption, identity verification, secret key exchange and other security issues is considered as a big challenge of establishing secure communication in computer networks. Resulting in the creation of artificial intelligence and its rapid expansion into various fields of network security. So, the general trend began to shift away from complex methods and algorithms with a mathematical foundation toward modern ones with a training and self-decision-making nature, which incorporate various concepts of artificial intelligence, such as the adoption of neural networks and deep learning.

The purpose of this study is to provide diverse works and research contributions that utilized various approaches of artificial intelligence to simulate the key security concepts that provide secure communication as shown in Fig. 1.

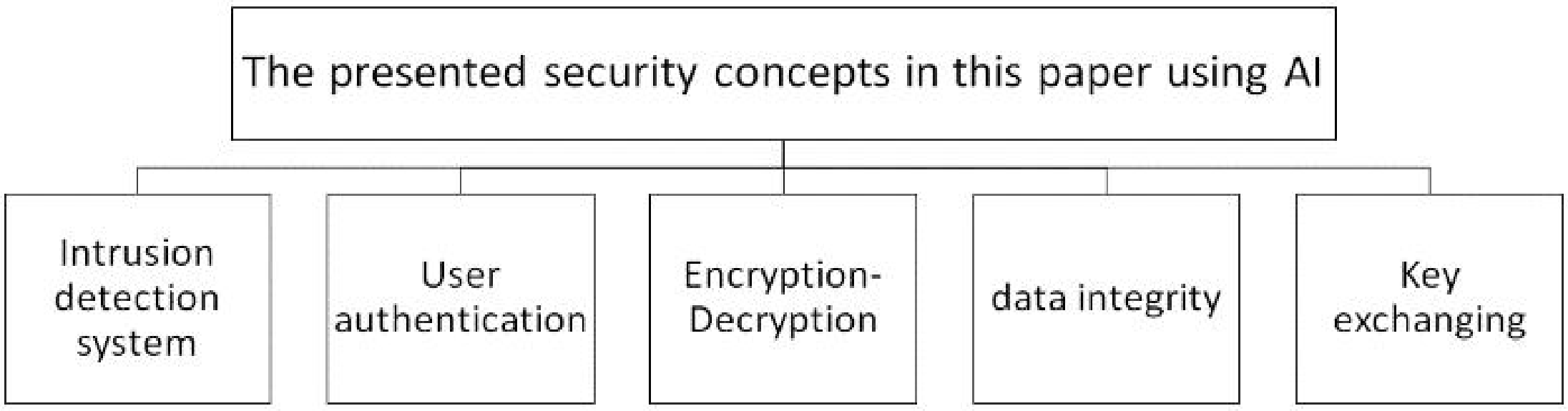

Fig. 1: Included security issues in our study

## 3.1 Intrusion detection system

An intrusion detection system based on supervised machine learning system combined with feature selection is proposed in [7] to classify network traffic whether it is malicious or not. The key role of feature selection component is to find most related attributes or features to recognize the instance to a particular class or group in addition to reduce data dimensionality. While, the role of learning algorithm component is to construct the necessary intelligence or knowledge utilizing the result obtained by the feature selection component as show in Fig.2.

The dataset is spilt into training and testing data, the model gets trained and establishes its intelligence based on the training data where each instance has the class it belongs to. Then, the model accuracy is evaluated by applying the learned intelligences to the testing dataset. Two different methods which are Wrapper method and filter method are used for feature selection process. Artificial neural network and Support Vector Machine SVM are employed for building the intelligence component of the model. The evaluation result demonstrates that the proposed model using Wrapper feature selection and ANN overcome all other models of filter method and SVM in network traffic classification with detection accuracy of 94.02% as illustrated in table 1.

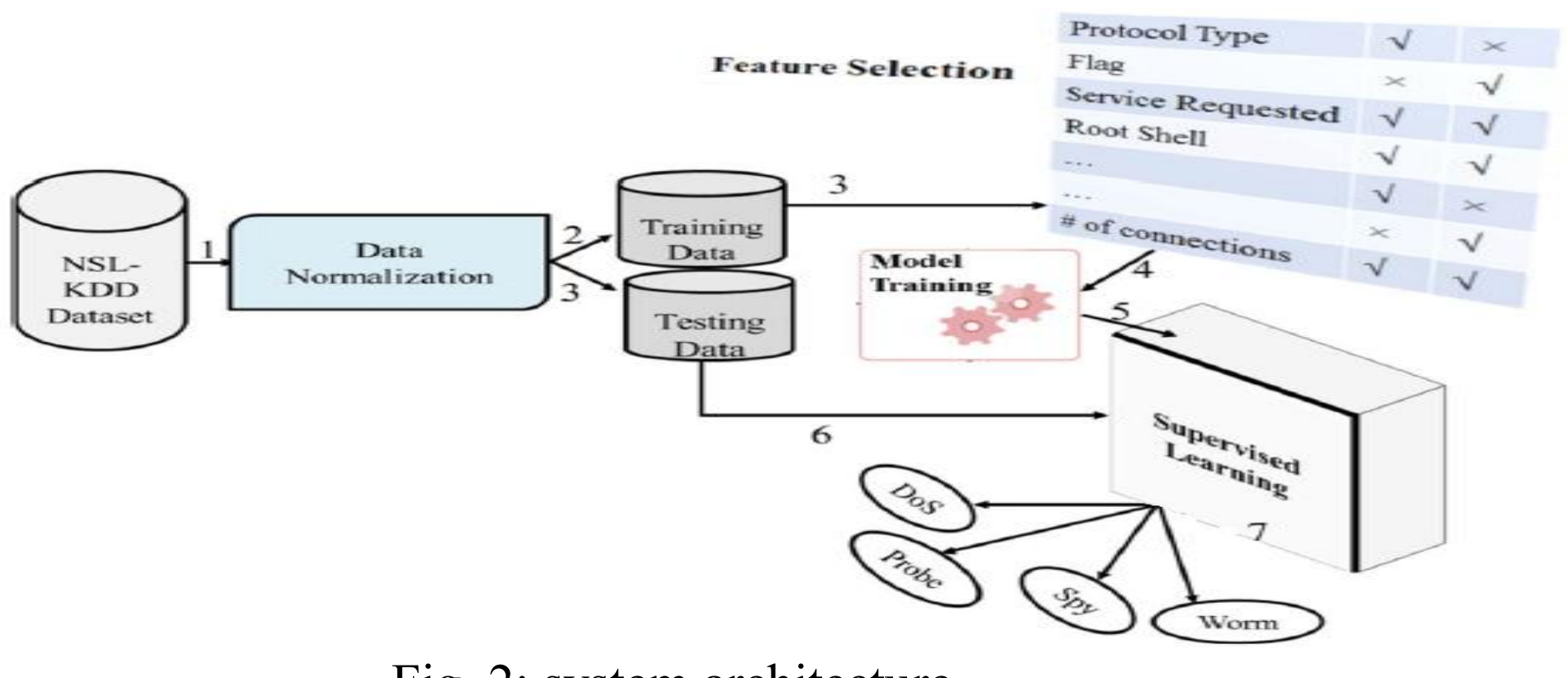

Fig. 2: system architecture

TABLE 1 RESULT OF CLASSIFICATION

| *Learning Type* | *Feature Selection* | *Input Features* | *Output Features* | *Detection Accuracy* |
|---|---|---|---|---|
| SVM | Filter | 41 | 35 | 82.34% |
| SVM | Wrapper | 41 | 17 | 81.78% |
| ANN | Filter | 41 | 35 | 83.68% |
| ANN | Wrapper | 41 | 17 | 94.02% |

## 3.2 Continuous User Authentication

A. T, Kiyani et al. [8] suggest a user authentication method based on deep neural network. Traditional methods are mostly used only once to validate the user's identity at the beginning of session. Instead, the suggested approach performs user authentication continuously based on keystroke dynamics, which checks user validity on each action based on the proposed robust recurrent confidence model (R-RCM). Moreover, Long Short Term Memory (LSTM) based recurrent neural network (RNN) is employed to handle the sequential nature of keystroke data. The patterns of keystroke can be represented as a sequential series comprising of key-down and key-press events. Formally, a keystroke sequence is a chronological organization of set of events representing a time series.

The suggested method decides the legitimacy of action based on the probability value of LSTM network which is utilize the previous actions as well. After that, R-RCM function receives the probability score from LSTM output layer and applies hyper parameters to determine if the user can continue using the system or should be locked out based on final threshold of R-RCM as shown in Fig. 3. Experimental results show that the deceiver users got detected just after performing 18% of keystroke events.

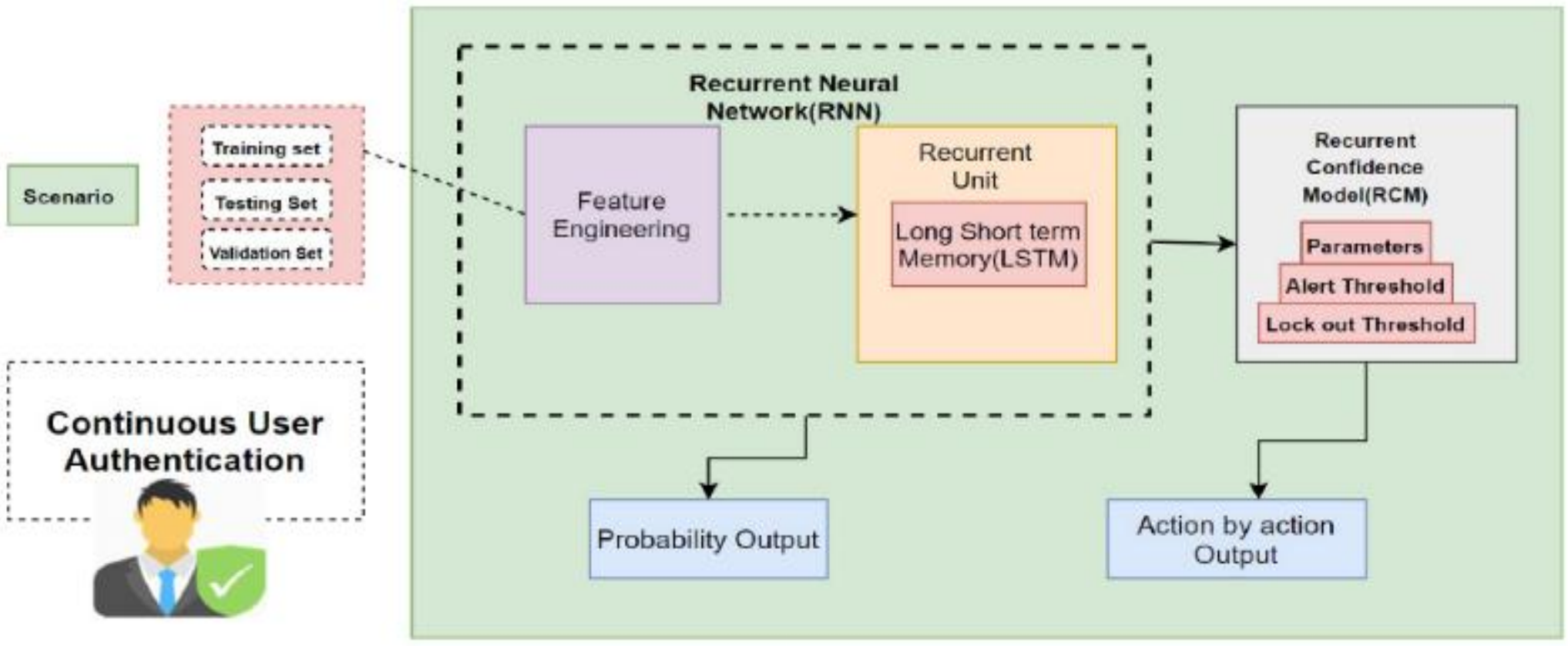

Fig. 3: System Architecture

## 3.3 On Demand Ciphering Engine

A lightweight highly secured On Demand Ciphering Engine (ODCE) based on artificial neural network is proposed in [9]. In order to increase the security level, the proposed method is characterized by keeping both the secret key and the ciphering engine secret, which are created by the user or the system administrator when transmission is initiated. These secrets are then exchanged by the involved parties. For achieving the principle of lightweight on demand ciphering engine, the proposed method is designed to be multi-option allowing the user to choose among various ciphering methods as shown in Fig. 4. In the first step, the user can choose between traditional ciphering methods or build his own new ciphering engine. If the user selects the traditional ciphering, another two sub-options are available: A traditional single ciphering method or a serial/parallel combination of various methods for better security level. On the other side, the second option enables the user to build his own cipher system using Feistel block cipher.

ANN is employed to transform the different ciphering engines forms into a unified circuit. The number of neurons in each layer depend essentially on the input block length. So, for each possible block length there is a certain ANN architecture. In the training process, ANN is fed with an input/output relation table for emulating the designed ciphering engine behavior. Neural network learning process continues until error rate reaches a certain threshold then the weights and biases of ANNs for both encryption and decryption are stored and prepared to be transmitted to the other side.

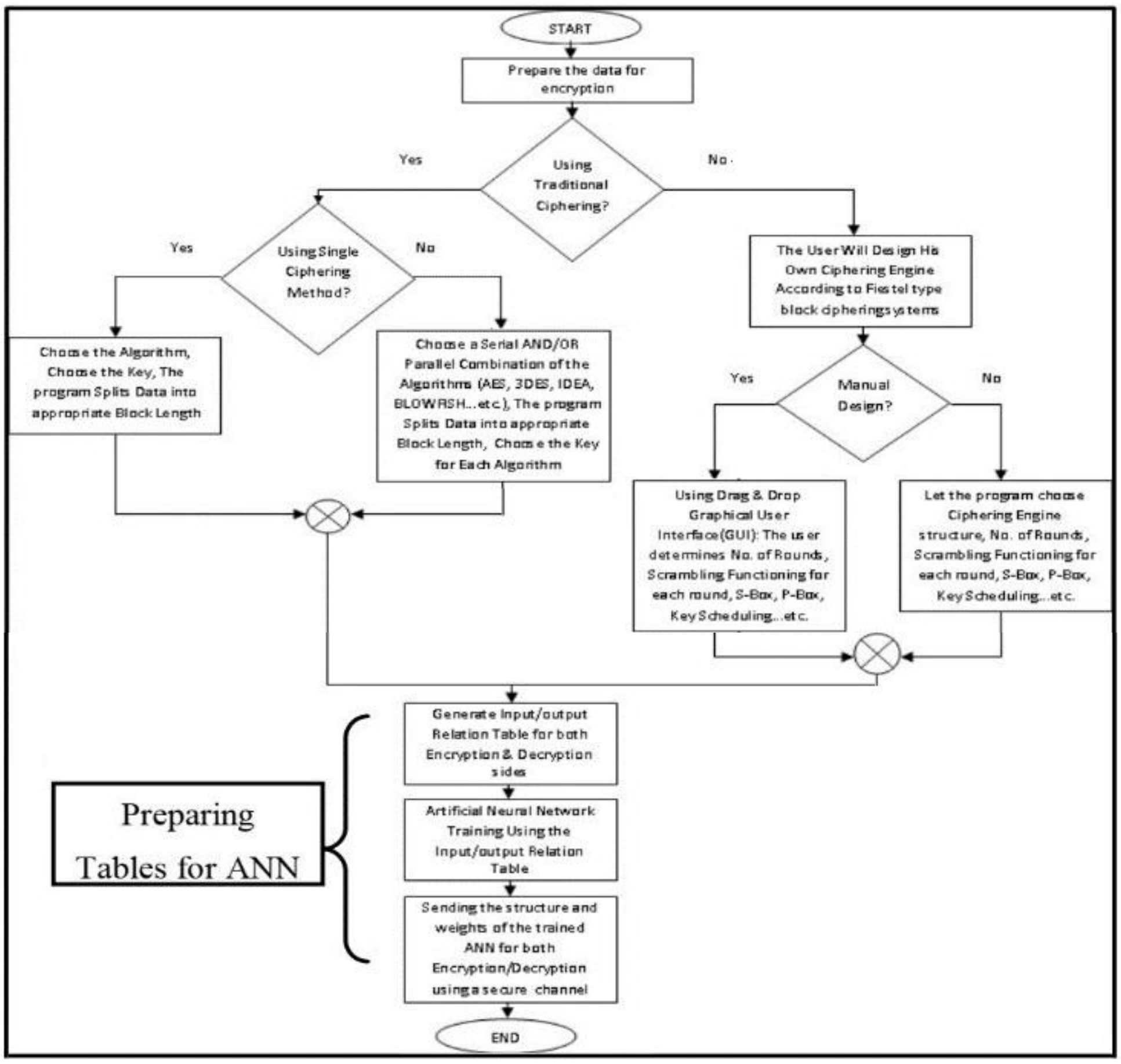

Fig. 4: Flow Chart of The Proposed ODCE Method

Experimental results show that, the adoption of ANN ODCE reduces the amount of transferred data to the weights and biases only which leads to the reduction of both size and transfer delay of file, as shown in Table 2.

TABLE 2 ODCE TRANSFER DELAY

| *Block size (Bit)* | *Traditional ODCE* | | *ANN ODCE* | |
|---|---|---|---|---|
| | *File Size of C++ Code* | *FTP Delay* | *ANN Weights and Biases Size* | *FTP Delay* |
| 8(SAES) | 15KB | 0.033s | 48B | 0.00024s |
| 64(3DES) | 200KB | 0.22s | 4KB | 0.0045s |
| 128(AES) | 100KB | 0.11s | 16KB | 0.018s |

## 3.4 Encryption Algorithm Based on Neural Network

An encryption technique to send data securely on communication network is proposed in [10] based on auto associative neural network which is trained by Hebbian rule. The main purpose of using auto associative neural network is to recognize the input pattern as two categories: known or unknown. An input pattern is recognized as known if the neural network produces the output precisely same as the input for that pattern. Hence, testing algorithm is used to check whether the network is distinguishing the given patterns or not. A private key, which is only known to the sender and the receiver, is represented by a matrix of size M*N, where M=2*N and N=4, 6, 8, 10 and so on. The even columns of the key matrix K contain 0s only and the odd columns can contain any combination of 1 and -1. The proposed algorithm has faster encryption and decryption speed. In case of using symmetric key, the researcher suggests that a trusted third party should be involved for preventing the key disclosing. Fig. 5 shows the training and testing algorithms. The encryption and decryption processes are illustrated in Fig. 6.

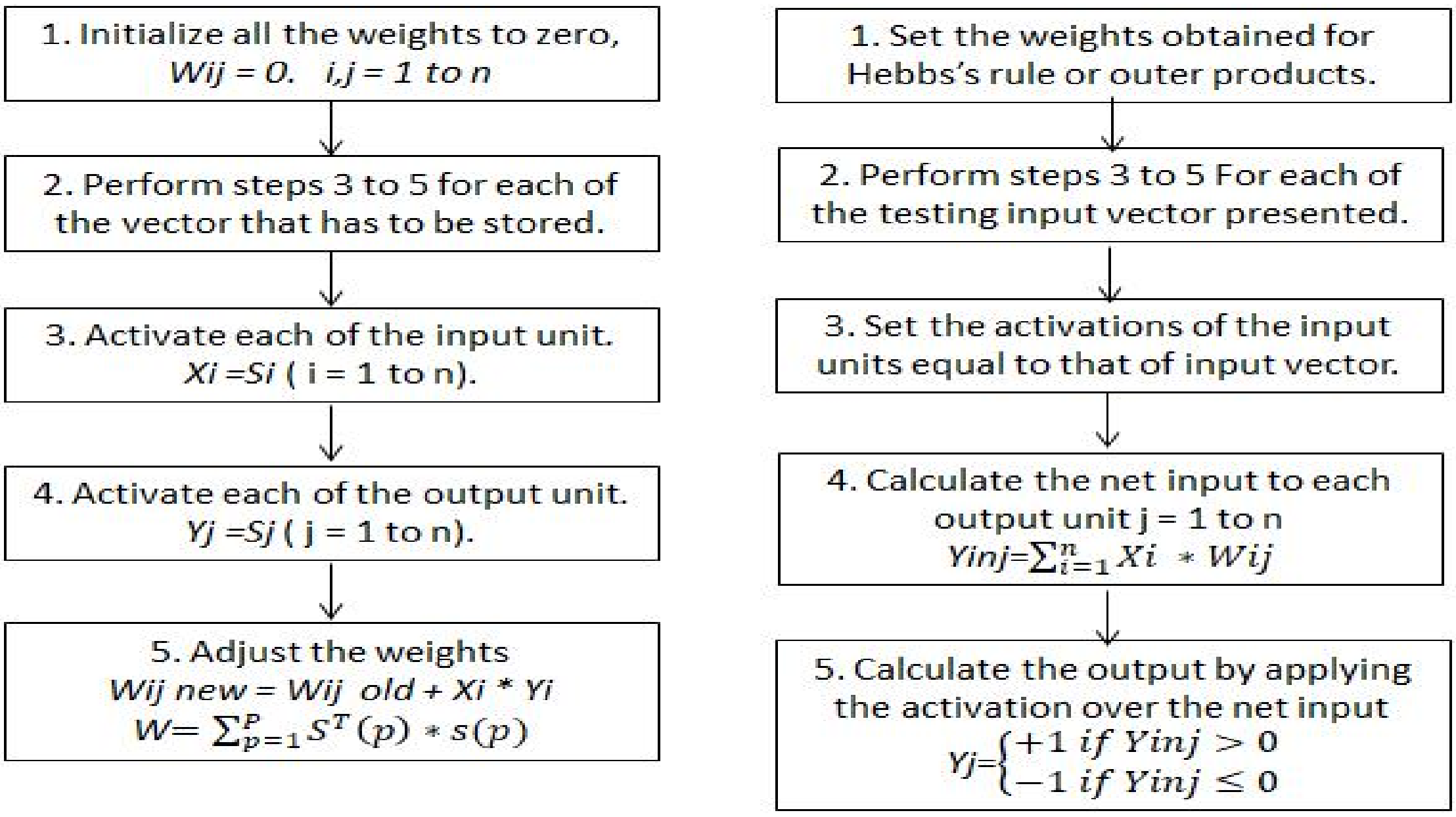

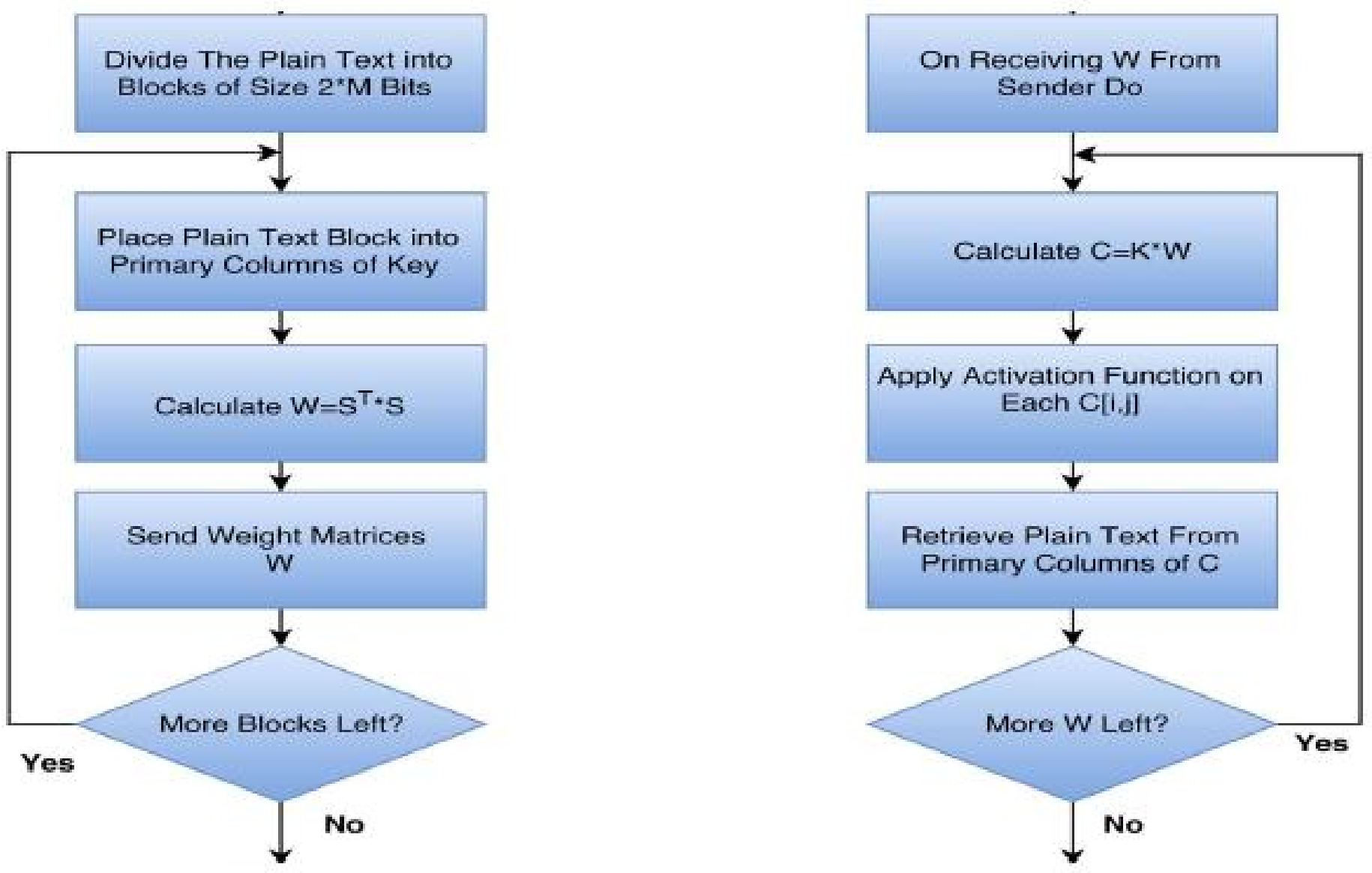

Fig. 6: Encryption and Decryption Algorithm

## 3.5 key exchange

In [11], an algorithm of key exchange using Perceptron neural network is proposed. In this network synaptic weights elements are synchronized into this network layer and in which the synaptic weights will be the secret keys in the Tree Parity Machines model. The Tree Parity Machines (TPM) model consists of an input vector X, an implicit Sigma $\delta$, a binding weight W between the input vectors and the hidden layer, and a set of activation functions that counts the resulting values τ, as illustrated in Fig. 7.

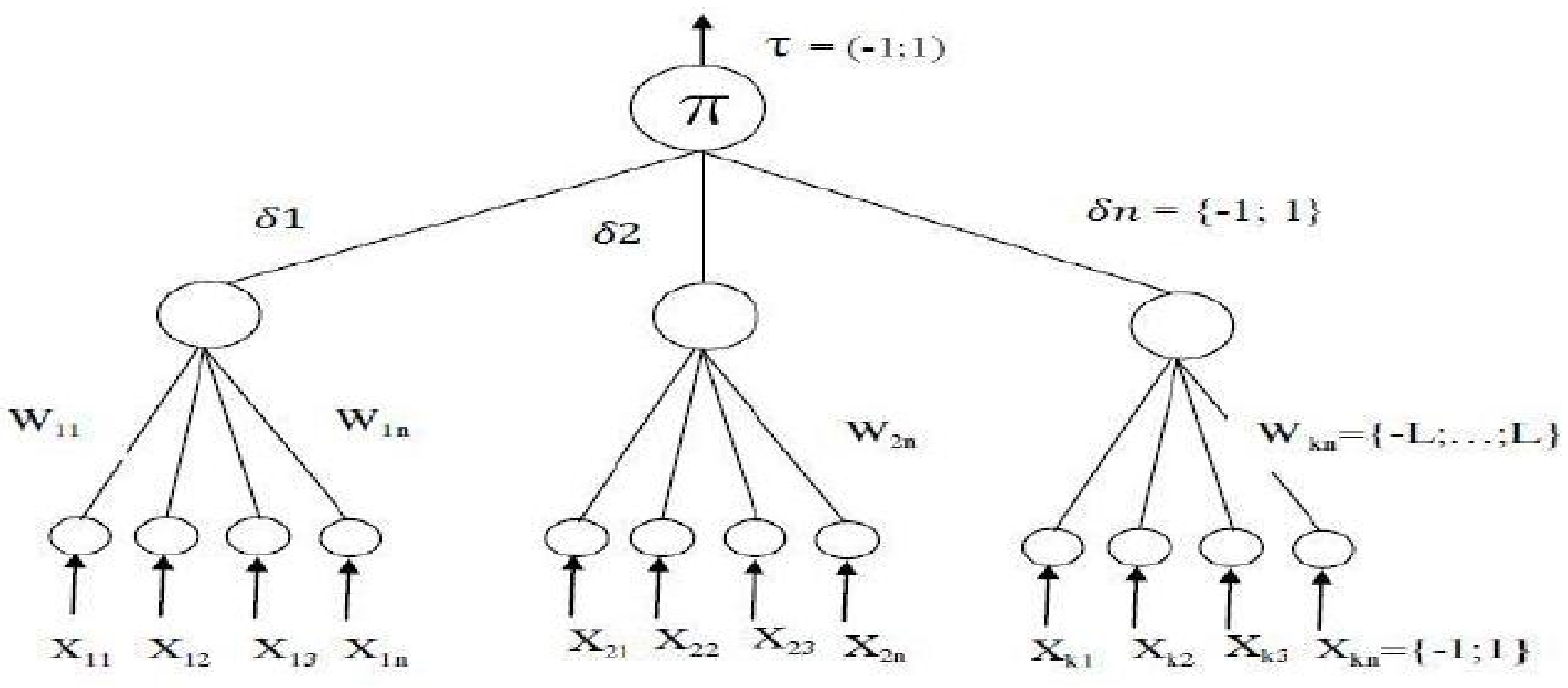

Fig. 7: Model of Tree Parity Machine

Fig. 8 is a descriptive chart for the presented algorithm, which can be summarized by the following steps:

1-setting up parameters for a neural network is done on both machine A and B.

2- the weights are generated randomly on each machine.

3-input vectors are set randomly on machines A and B.

4- Calculate the output vectors and exchange between machines A and B.

5- If the output vectors on both machines are the same means τA = τB, then it is necessary to continue to step 6. Otherwise, step 3 should be repeated.

6-The corresponding weights per machine are updated using the Hebbian training rule. If the weights are the same, it is necessary to continue to 7 if not, step 3 should be repeated.

7-Finally, the weights of the neural network are the same for both machines. In addition, these weights are used to generate the secret key .

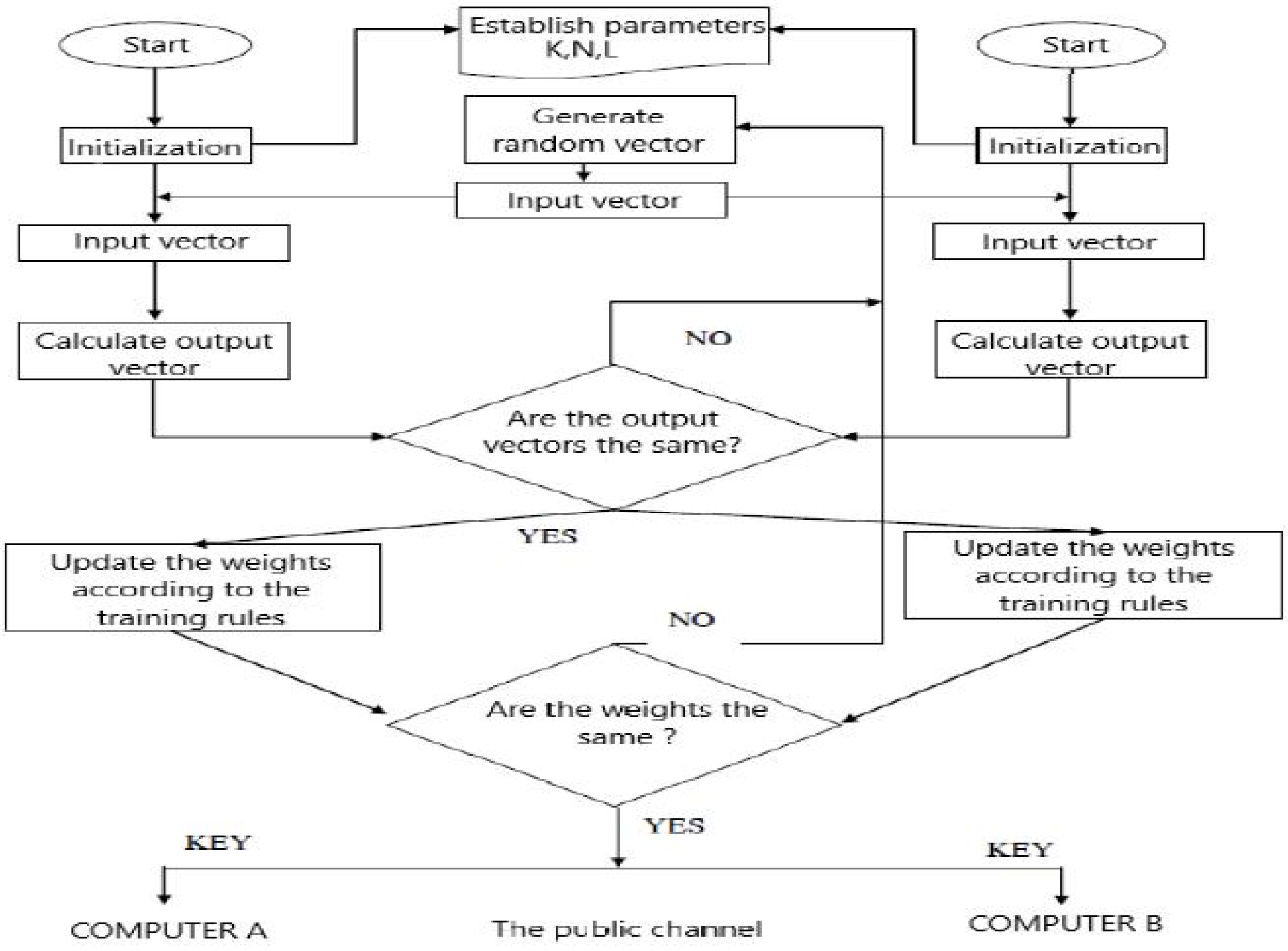

Fig. 8: key exchange algorithm

## 3.6 Integrity- Hash function

M. Turčaník utilize a RNN to present a new schema for hash function of 128 bits [12]. A recurrent neural network (RNN) is composed from three layers of neurons and synapses between them.as shown in Fig. 9. The Proposed hash function algorithm is clarified as flowchart in Fig10 which can be summarized by the following steps:

1- Transformation of message into sequence of bits. This sequence of bits is split into 512 bits long blocks.

2- generation of initial message digest value which is used as input for hash function calculation (128 bits).This operation is realized only in first step of calculation because we don't have any hash value yet. After realization of first step hash value is used in the next step of calculation together with next 512 bits block of the massage.

3-message which is divided onto blocks is sent to the input of the recurrent neural network. The hash function computation is finished when all blocks of the message is processed. Last value on the output of the recurrent neural network is final hash function value.

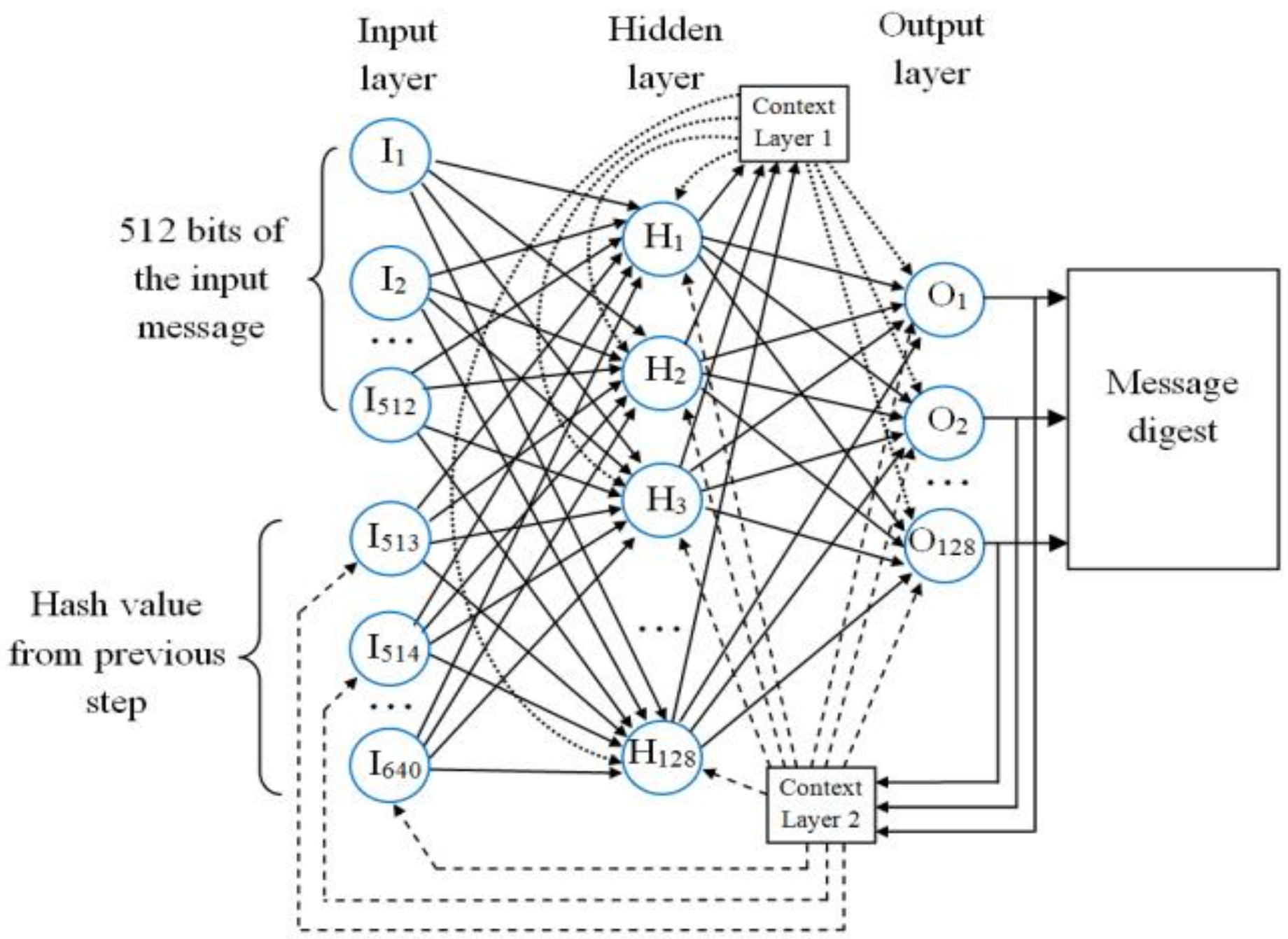

Fig. 9: Structure of the RNN for hash function generation

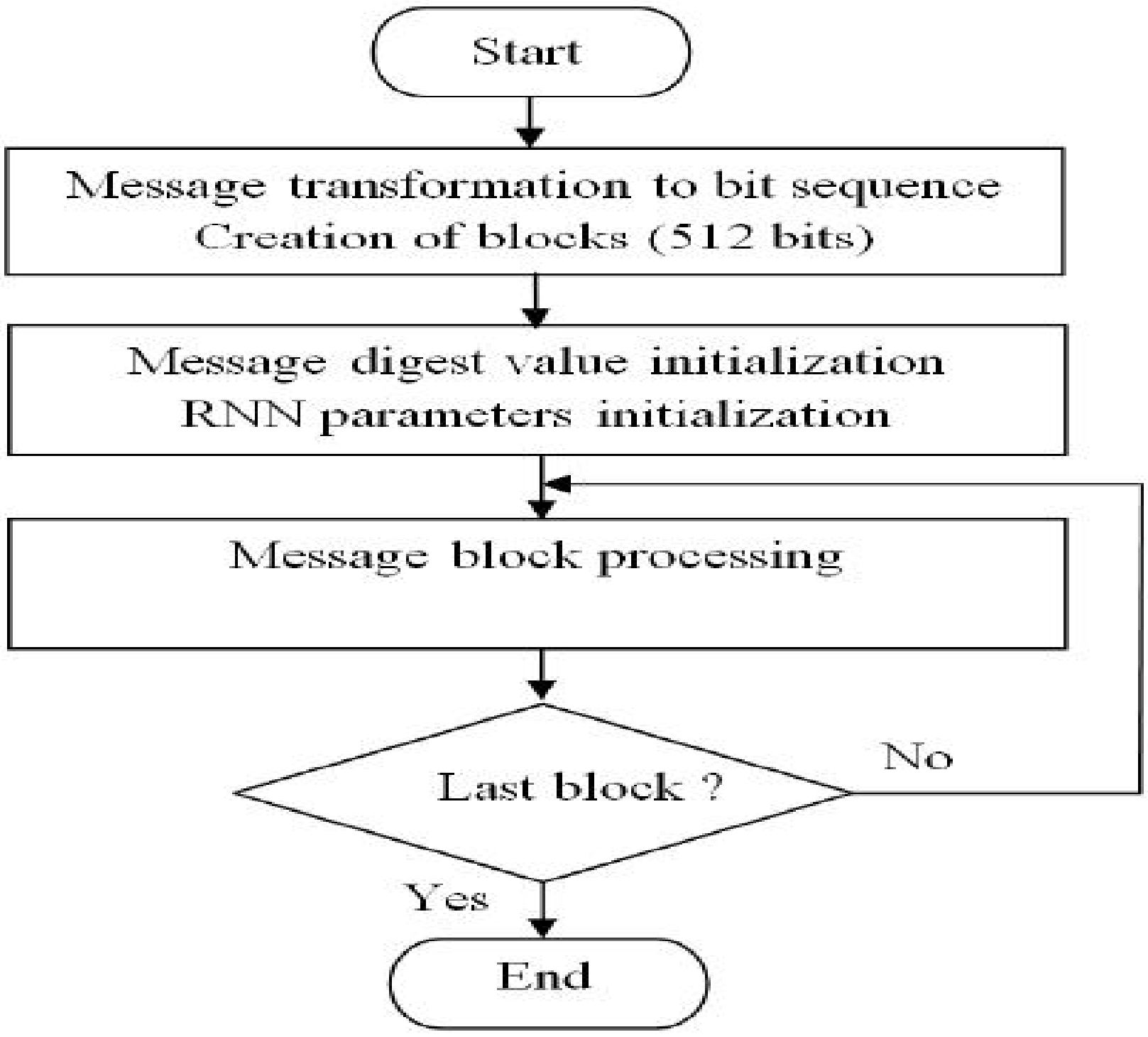

Fig. 10: The algorithm of the hash function calculation

## 3.7 Fast decryption method

A. Haggag et al. in [13], propose a fast approach for decrypting small encrypted communications of ten letters using the trained Feed-Forward Artificial Neural Network methodology (FF-ANN). This method can be applied to a huge message that requires the use of mathematics for encryption and decryption. Feathers are supplied to FF-ANN, which is trained with 50 encrypted messages. One of the encryption methods employed is asymmetric key elliptic curve cryptography (ECC). The proposed method achieves an excellent result in that decryption takes a short time, approximately one minute, while it takes approximately ten minutes and approximately twenty minutes for small and medium messages, respectively, and its benefit is obvious when the message contains a large number of letters as shown in Fig. 11.

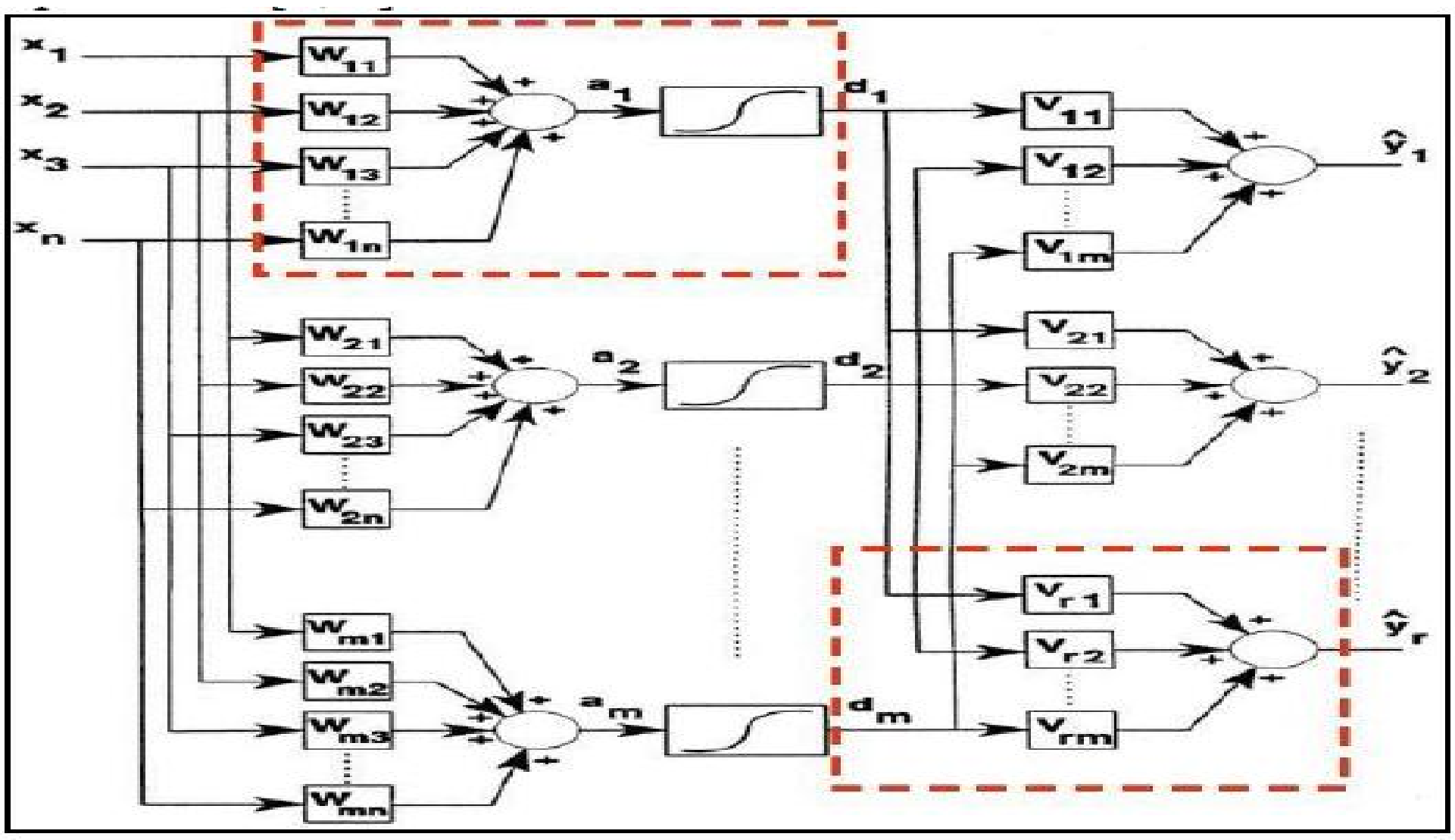

Fig. 11: Feed-forward and back-propagation operations

## 3.8 CNN based encryption method

In [14] the authors presents an implementation of Convolutional Neural Network (CNN) symmetric encryption. Since the input which is a plaintext is store as a character (8-bit element), is shaping 3D block size (8 or 16) as shown in Fig. 12.

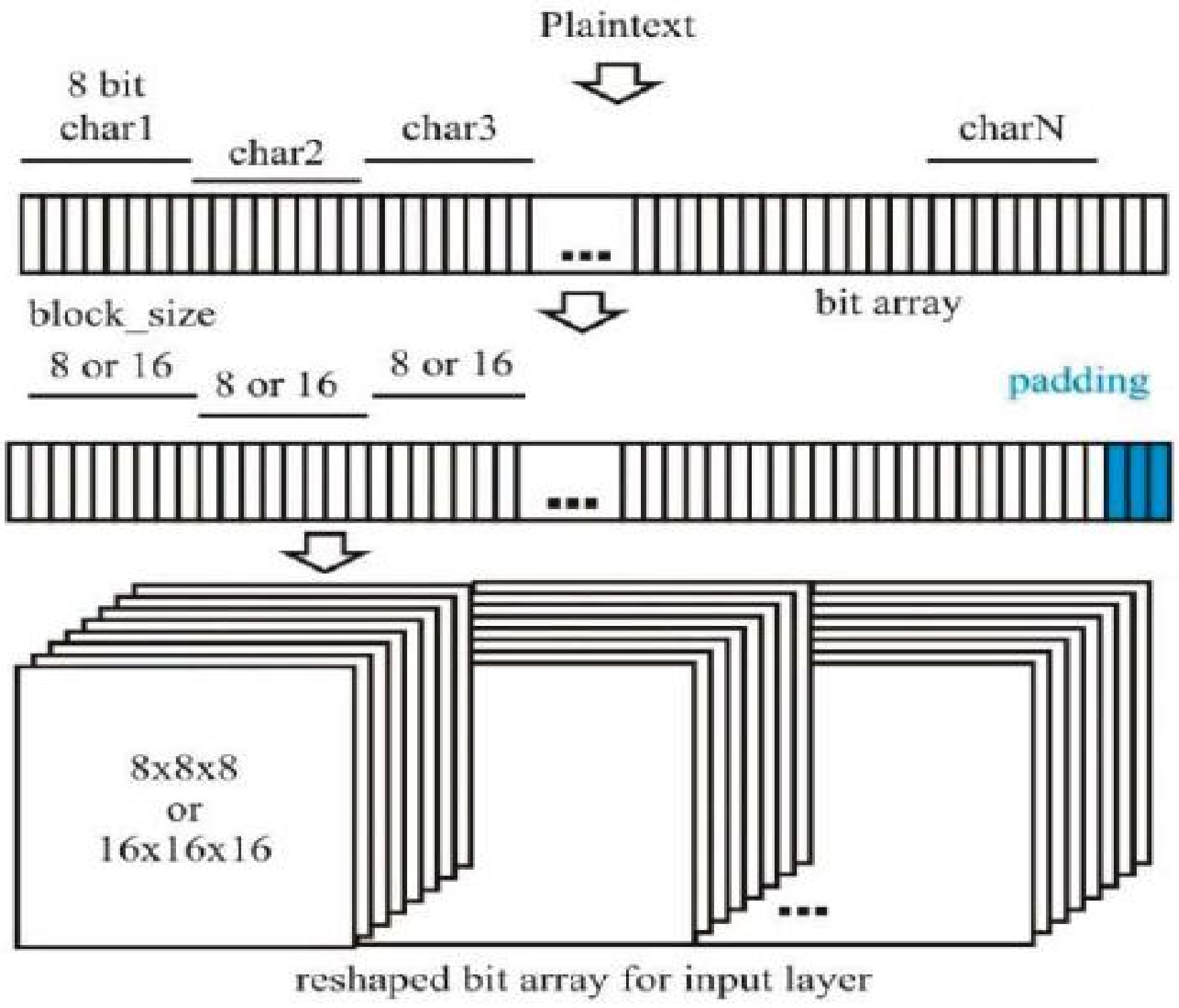

Fig. 12: Data shaping block size 8 or 16.

The crypto key is specified as a binary cube with binary values randomly generated and can be produced with a user password. The cube size is obtained from the input parameter block size. Final crypto key is saved to the file in a flattened one dimensional character form. The crypto key is used as a convolutional kernel for CNN. Keras of 4 layers (input layer, conv3D , flatten layer and output layer) is used to construct neural network as shown in Fig. 13.

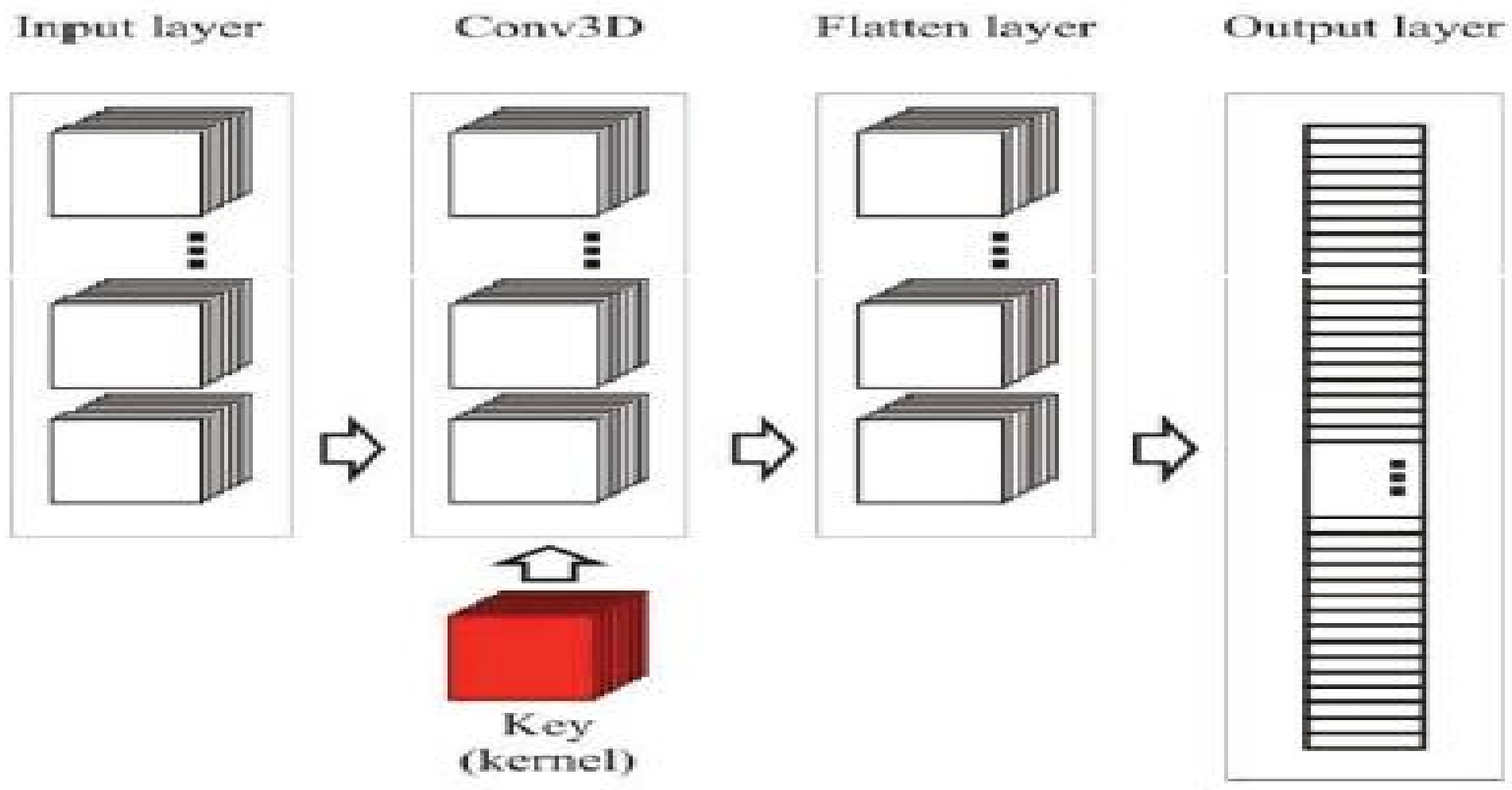

Fig. 13: Structure of CNN for encryption

## 3.9 Neural network based user authentication

In [15], M. Dahiya propose user authentication mechanism using neural networks that the network be trained to store (stored in the form of network parameter) passwords rather than relying on a validation table. The user ID and password are encoded in the proposed system to n-bit binary numbers that are utilized to train the BPNN neural network. The neural network is still iterating until the user ID and password pairs are remembered. Fig. 14 depicts the system's schematic. Fig. 14 clarifies the architecture of the proposed system. The designed system evaluated using of , A PC with Intel-core i5 processor, 2.40GHz and 8GB RAM, for thirty users, and The total time for the registration of 30 users was about 0.0005 sec.

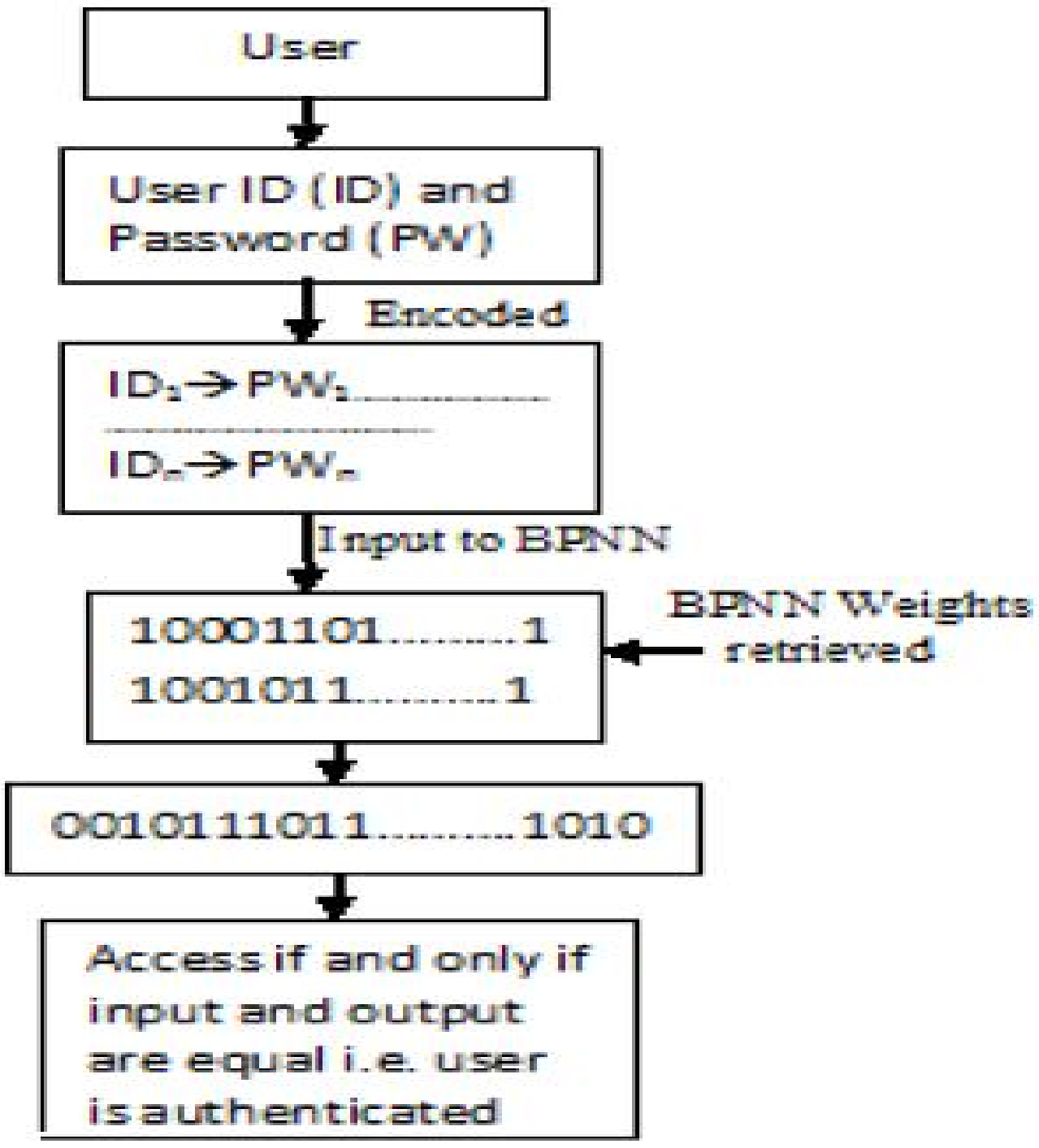

Fig. 14: User Authentication using BPNN

## 3.10 Neural network based cryptology system

A cryptology system is proposed by [16], which mainly is based on both of a pseudo random number and neural network. Fig. 15 shows the overall architecture of the presented system that is consisting of six steps for (encryption/decryption) which are summarized as follows:

step 1: Represents plaintext input.

step 2: Neural based pseudo-random numbers are generated via what is call modified subtract with borrow.

step 3: Represents non-linear encryption, and can be summarize by converting plain text to ASCII codes, converting ASCII codes to binary codes, mixing digit numbers of each string's binary codes and mixing digit numbers of all strings' binary codes, in order to create neural-based pseudo-random numbers.

step 4: involves developing the neural network architecture for pseudo-random numbers. The modified subtract with borrow algorithm generates random integers that are utilized as input values, initial weight, bias values, and the neuron count of hidden layers. Without training, the output values of the network are evaluated, and the enhanced random numbers are used to construct the cryptosystem.

step 5: Sending neural network topology and chipper text, are accomplished in this stage.

step 6: Represents the simulation of neural network and decryption process.

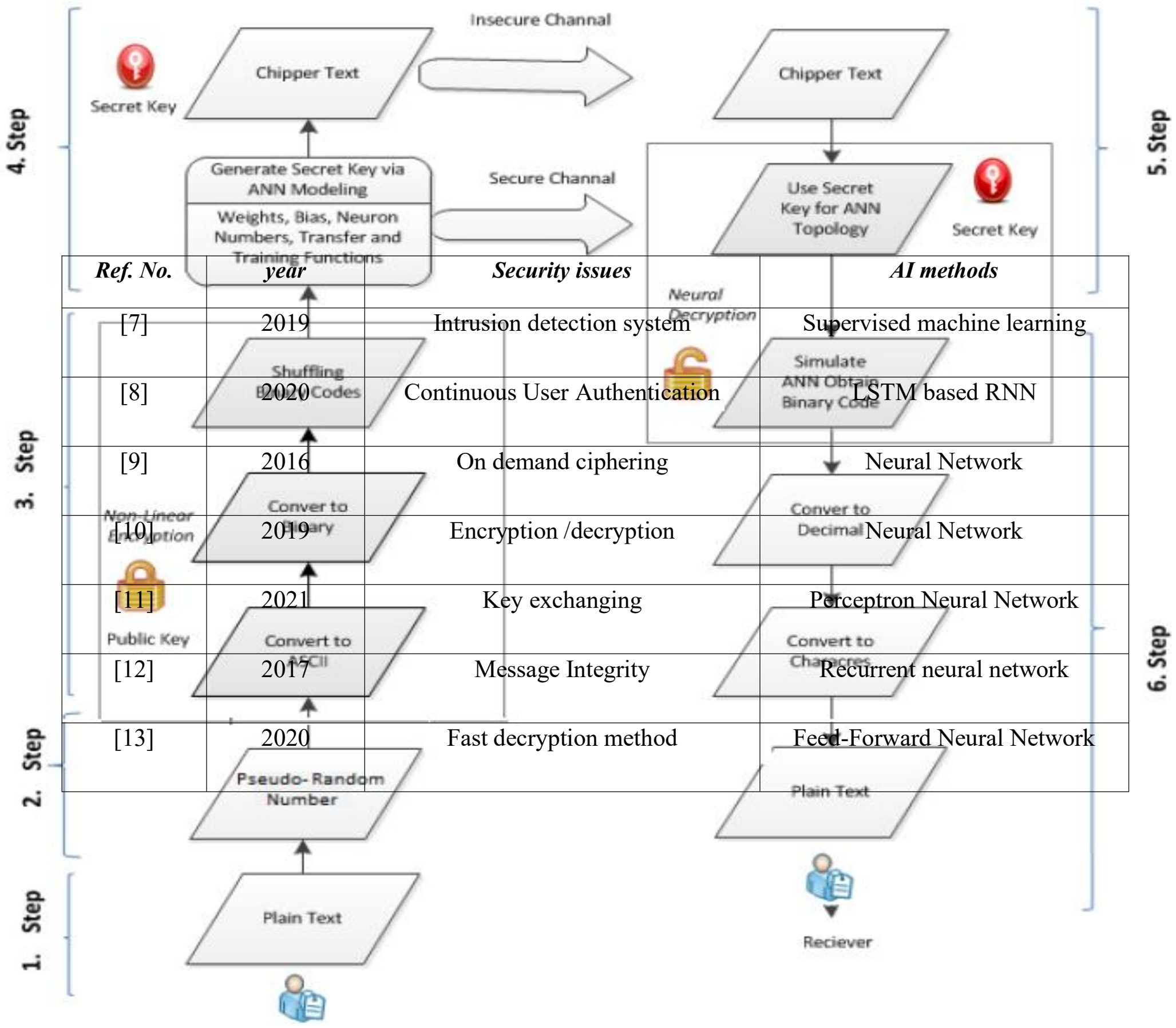

Fig. 15: Flowchart of proposed neural cryptosystem

Finally, table III briefly summarizes the researches of artificial intelligence recently employed in security issues which are reviewed in this study.

Table 3 Summary of reviewed works

| *Ref. No.* | *year* | *Security issues* | *AI methods* |
|---|---|---|---|
| [7] | 2019 | Intrusion detection system | Supervised machine learning |
| [8] | 2020 | Continuous User Authentication | LSTM based RNN |
| [9] | 2016 | On demand ciphering | Neural Network |
| [10] | 2019 | Encryption /decryption | Neural Network |
| [11] | 2021 | Key exchanging | Perceptron Neural Network |
| [12] | 2017 | Message Integrity | Recurrent neural network |
| [13] | 2020 | Fast decryption method | Feed-Forward Neural Network |
| [14] | 2019 | Encryption method | Convolutional Neural Network |
| [15] | 2016 | User Authentication | Back propagation Neural Network |
| [16] | 2014 | Cryptology system | Neural Network |

## 4. Conclusion

The fundamental goal of using artificial intelligence in network security is to move away from traditional security strategies that adopt mathematical and statistical methods in their approach, which are take some time for their implementation as well as it have some restrictions, toward innovative and self-tuning methods in which human intervention is limited and adopt the principle of automation in its structure resulting more secure system at lower cost and time. This is what has been highlighted by this study including various artificial intelligence based methods and strategies to accomplish most of the aspects of secure communication in terms of encryption/decryption, authentication, key exchanging, data integrity and malware detection.